\documentclass[article]{aa}  

\usepackage{graphicx}
\usepackage{subfig}
\usepackage{txfonts}

\begin{document}

   \title{First detection of cyanamide (NH$_2$CN) towards solar-type protostars}

   \author{A. Coutens \inst{1} 
            \and E. R. Willis \inst{2}
         \and R. T. Garrod \inst{2}
         \and H. S. P. M\"uller \inst{3}
         \and T. L. Bourke \inst{4}
         \and H. Calcutt \inst{5}
         \and M. N. Drozdovskaya \inst{6}
         \and J. K. J\o rgensen \inst{5}
         \and N. F. W. Ligterink \inst{7}
         \and M. V. Persson \inst{8}
         \and G. St\'ephan \inst{2}
         \and M.~H.~D. van der Wiel \inst{9}
         \and E. F. van Dishoeck \inst{7,10}
         \and S. F. Wampfler \inst{6}
          }

   \institute{Laboratoire d'Astrophysique de Bordeaux, Univ. Bordeaux, CNRS, B18N, all\'ee Geoffroy Saint-Hilaire, 33615 Pessac, France \\
              \email{audrey.coutens@u-bordeaux.fr}
	\and Departments of Chemistry and Astronomy, University of Virginia, Charlottesville, VA 22904, USA
	\and I. Physikalisches Institut, Universit\"at zu K\"oln, Z\"ulpicher Str. 77, 50937 K\"oln, Germany
	\and SKA Organization, Jodrell Bank Observatory, Lower Withington, Macclesfield, Cheshire SK11 9DL, UK
         \and Centre for Star and Planet Formation, Niels Bohr Institute and Natural History Museum of Denmark, University of Copenhagen, \O ster Voldgade 5-7, DK-1350 Copenhagen K, Denmark
         \and Center for Space and Habitability, Universit\"at Bern, Sidlerstrasse 5, 3012 Bern, Switzerland
	\and Leiden Observatory, Leiden University, PO Box 9513, 2300 RA Leiden, The Netherlands        
         \and 
         Department of Space, Earth and Environment, Chalmers University of Technology, Onsala Space Observatory, 439 92, Onsala, Sweden
         \and ASTRON Netherlands Institute for Radio Astronomy, PO Box 2, 7990 AA Dwingeloo, The Netherlands
         \and Max-Planck Institut f\"ur Extraterrestrische Physik (MPE), Giessenbachstr. 1, 85748 Garching, Germany
             }

   \date{Received xxx; accepted xxx}

 
  \abstract
  {Searches for the prebiotically-relevant cyanamide (NH$_2$CN) towards solar-type protostars have not been reported in the literature. 
  We here present the first detection of this species in the warm gas surrounding two solar-type protostars, using data from the Atacama Large Millimeter/Submillimeter Array Protostellar Interferometric Line Survey (PILS) of IRAS 16293--2422 B and observations from the IRAM Plateau de Bure Interferometer of NGC1333~IRAS2A. We furthermore detect the deuterated and $^{13}$C isotopologues of NH$_2$CN towards IRAS 16293--2422 B. This is the first detection of NHDCN in the interstellar medium. Based on a local thermodynamic equilibrium analysis, we find that the deuteration of cyanamide ($\sim$ 1.7\%) is similar to that of formamide (NH$_2$CHO), which may suggest that these two molecules share NH$_2$ as a common precursor. The NH$_2$CN/NH$_2$CHO abundance ratio is about 0.2 for IRAS 16293--2422 B and 0.02 for IRAS2A, which is comparable to the range of values found for Sgr B2. We explored the possible formation of NH$_2$CN on grains through the NH$_2$ + CN reaction using the chemical model MAGICKAL. Grain-surface chemistry appears capable of reproducing the gas-phase abundance of NH$_2$CN with the correct choice of physical parameters. 
}
 

   \maketitle
%

\section{Introduction}

Cyanamide (NH$_2$CN) is one of the rare interstellar molecules that contain two atoms of nitrogen. 
This species is thought to be relevant for prebiotic chemistry, since, in liquid water, it may convert into urea, an important molecule in biological processes  \citep{Kilpatrick1947}.
Its isomer carbodiimide (HNCNH) can be formed from NH$_2$CN in photochemically and thermally induced reactions in interstellar ice analogues \citep{Duvernay2005}. Molecules with the carbodiimide moiety (--NCN--) find use in various biological processes, among
which the assembly of amino acids into peptides (see \citealt{Williams1981} for an overview). 
Although this molecule is detected in other galaxies, such as NGC 253 and M82 \citep{Martin2006,Aladro2011}, only two detections are mentioned in our Galaxy: the massive star-forming regions Sgr B2 \citep{Turner1975} and Orion KL \citep{White2003}.

Molecules formed early during the star formation process may be incorporated into comets or asteroids and delivered to planets during heavy bombardment periods, similar to that experienced by the young Earth \citep{Raymond2004}.
Here, we report the detection of cyanamide towards two solar-type protostars, IRAS~16293--2422 (hereafter IRAS16293) and NGC1333 IRAS2A (hereafter IRAS2A).
These two low-mass protostars are known to harbor a very rich chemistry in their inner regions \citep[e.g.,][]{Bottinelli2004,Jorgensen2005}. This can be explained by the thermal desorption of the numerous and complex species formed in the icy mantles of the grains. The detection presented in this work complements the list of molecules of prebiotic interest such as glycolaldehyde, formamide and methyl isocyanate discovered in low-mass protostars \citep{Jorgensen2012,Kahane2013,Maury2014,Coutens2015,Ligterink2017,Martin2017}. We also detect the deuterated form of cyanamide, NHDCN, for the first time in the interstellar medium.


\section{Observations}
\label{sect_obs}

To search for cyanamide, we used data obtained with the Atacama Large Millimeter/Submillimeter Array (ALMA) for IRAS16293 and with the IRAM Plateau de Bure Interferometer (PdBI) for IRAS2A.

The ALMA data are part of the PILS ("Protostellar Interferometric Line Survey") program$\footnote{\url{http://youngstars.nbi.dk/PILS/}}$, a large spectral survey of IRAS16293 observed in Cycle 2 between 329.1 and 362.9 GHz at a spatial resolution of about 0.5$\arcsec$ and a spectral resolution of $\sim$0.2 km s$^{-1}$. The observations and their reduction are presented in \citet{Jorgensen2016}. The data reach a sensitivity of about 4--5 mJy beam$^{-1}$ km s$^{-1}$.

The PdBI data of the low-mass protostar IRAS2A were obtained with the WIDEX correlator in the framework of several programs (V010, V05B, W00A, and X060). The data reduction of each dataset is described in \citet{Coutens2014,Coutens2015} and \citet{Persson2014}.
They cover the spectral ranges 223.5--227.1, 240.2--243.8, and 315.5--319.1 GHz with a spectral resolution of 1.95 MHz ($\sim$1.8--2.6 km s$^{-1}$). The angular resolution is about 1.2$\arcsec$ $\times$ 1.0$\arcsec$ at 225 GHz, 1.4$\arcsec$ $\times$ 1.0$\arcsec$ at 242 GHz, and 0.9$\arcsec$\,$\times$\,0.8$\arcsec$ at 317 GHz. The rms are $\sim$5--6\,mJy\,beam$^{-1}$ km s$^{-1}$ or lower. 

The spectroscopic data used here for cyanamide and its $^{13}$C isotopologue come from the spectroscopic catalogs Jet Propulsion Laboratory (JPL, \citealt{Pickett1998,Read1986}) and Cologne Database for Molecular Spectroscopy (CDMS, \citealt{Muller2001,Muller2005,Krasnicki2011}). 
The spectroscopy of NHDCN was studied by \citet{Kisiel2013}.
Carbodiimide spectroscopic information comes from the CDMS \citep{Birk1989,Jabs1997}.

\section{Results}
\label{sect_results}

The CASSIS$\footnote{CASSIS has been developed by IRAP-UPS/CNRS (\url{http://cassis.irap.omp.eu/}).}$ software was used to search for and identify the lines of NH$_2$CN and its isotopologues towards IRAS16293 and IRAS2A. Synthetic spectra were produced and compared with the observations to identify the lines. Potential blending with other species from the CDMS or JPL catalogs was checked. Column densities were determined assuming local thermodynamic equilibrium (LTE), which is reasonable for the inner regions of low-mass protostellar envelopes owing to their very high densities ($\gtrsim$\,10$^{10}$ cm$^{-3}$, \citealt{Jorgensen2016}).

\subsection{Analyses of cyanamide and carbodiimide}

\begin{figure}[t!]
\begin{center}
\includegraphics[width=9cm]{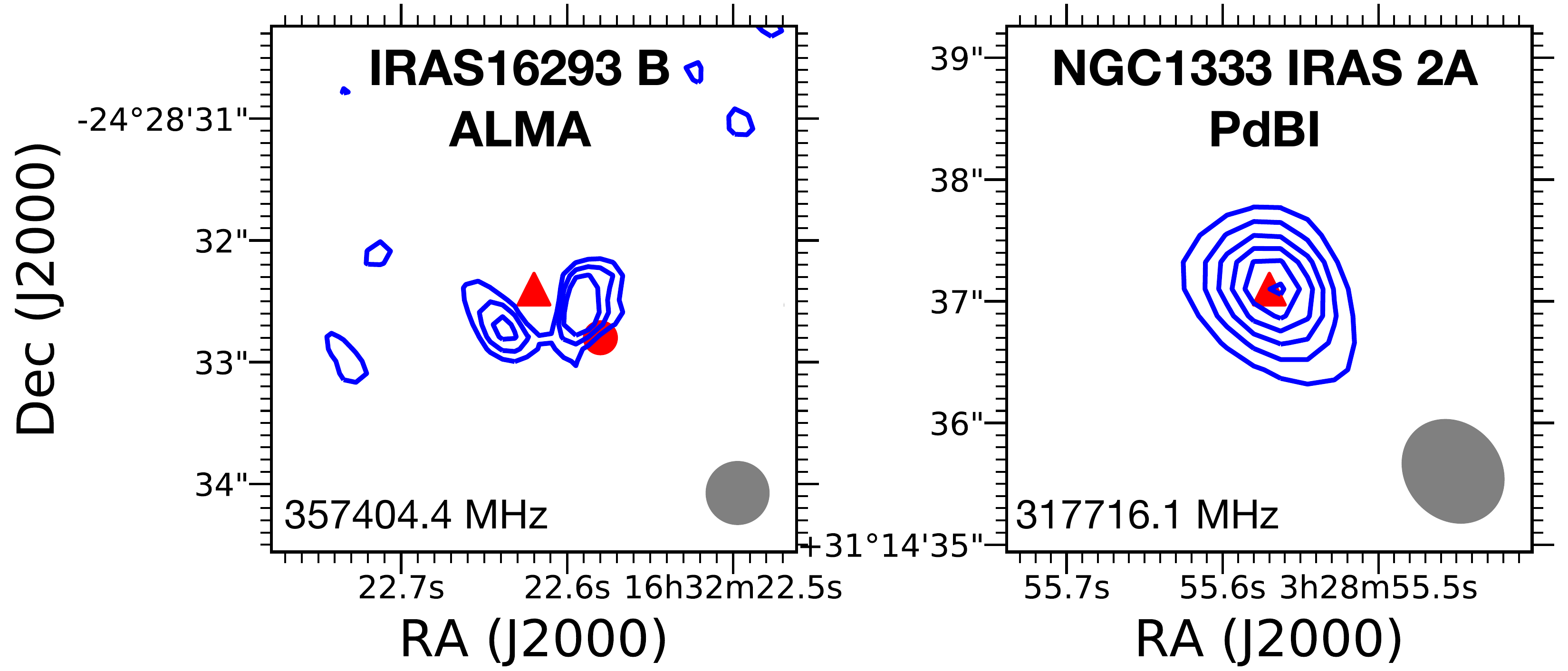}
\caption{Integrated intensity maps of two transitions of NH$_2$CN detected towards IRAS16293 B (left panel, from 3 to 5$\sigma$ by step of 1$\sigma$) and IRAS2A (right panel, from 5 to 25$\sigma$ by step of 5$\sigma$). The position of the continuum peak position is indicated with a red triangle, while the position analyzed for IRAS16293 B (full-beam offset) is indicated with a red circle. The beam size is shown in grey in the right hand lower corner.}
\label{maps_NH2CN}
\end{center}
\end{figure}

\begin{figure*}[ht!]
\begin{center}
\includegraphics[width=18cm]{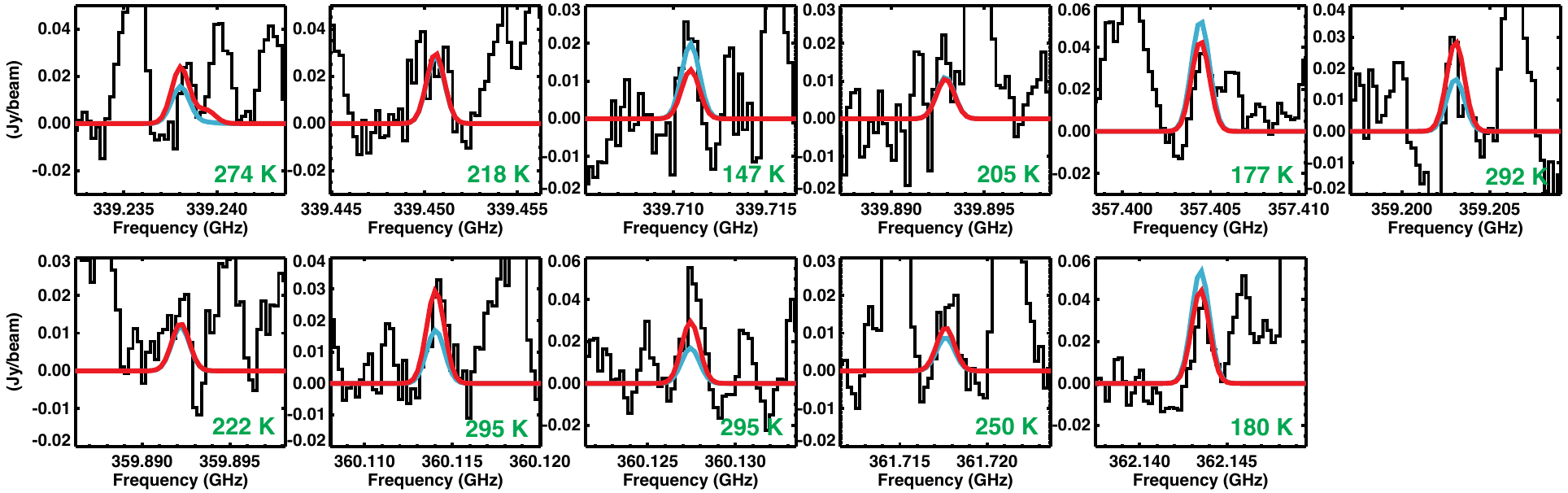}
\caption{Unblended lines of NH$_2$CN detected with ALMA towards IRAS16293 B at the full-beam offset position. The best-fit model for $T_{\rm ex}$ = 300 K and 100 K are in red and blue respectively. The $E_{\rm up}$ values are indicated in green in the bottom right corner of each panel.}
\label{figure_detection_iras16293}
\end{center}
\end{figure*}

For the binary IRAS16293, eleven unblended lines of the main cyanamide isotopologue are detected towards source B (see Figure \ref{figure_detection_iras16293}) at the full-beam offset position analyzed in previous studies \citep{Coutens2016,Lykke2017,Ligterink2017}. No clear detection could be obtained towards IRAS16293 A, where the lines are broader ($\geq$ 2 km\,s$^{-1}$). 
Maps (see Figure \ref{maps_NH2CN}) indicate that the emission of this species arises from the warm inner regions around the B component, similarly to other complex organic molecules \citep{Baryshev2015,Jorgensen2016}. This species also appears to be strongly affected by absorption against the strong continuum similarly to formamide \citep[][see their Figure 1]{Coutens2016}. The deep absorptions at the continuum peak and half-beam offset positions are  clearly seen for all the lines (see Figure \ref{figure_absorption}). Although a LTE model with a lower excitation temperature of 100\,K is in relatively good agreement with the observations (see Figure \ref{figure_detection_iras16293}), a temperature of 300 K is a more appropriate fit. This is consistent with the temperature derived for other species with high binding energies such as glycolaldehyde, ethylene glycol, and formamide \citep[see discussion in][]{Jorgensen2017}. 
At the full-beam offset position, a column density of $\geq$\,7\,$\times$\,10$^{13}$ cm$^{-2}$ is derived for an excitation temperature of 300 K ($\geq$\,5\,$\times$\,10$^{13}$ cm$^{-2}$ for 100\,K) and a source size of 0.5$\arcsec$. It should be noted that this column density can only be considered as a lower limit, because of the absorption components that could lower the emission contribution of the line profile. The higher value of the column density of NH$_2$CN is confirmed by the analysis of the $^{13}$C isotopologue (see Section \ref{sect_isotopologues}).

\begin{figure}[ht!]
\begin{center}
\includegraphics[width=9cm]{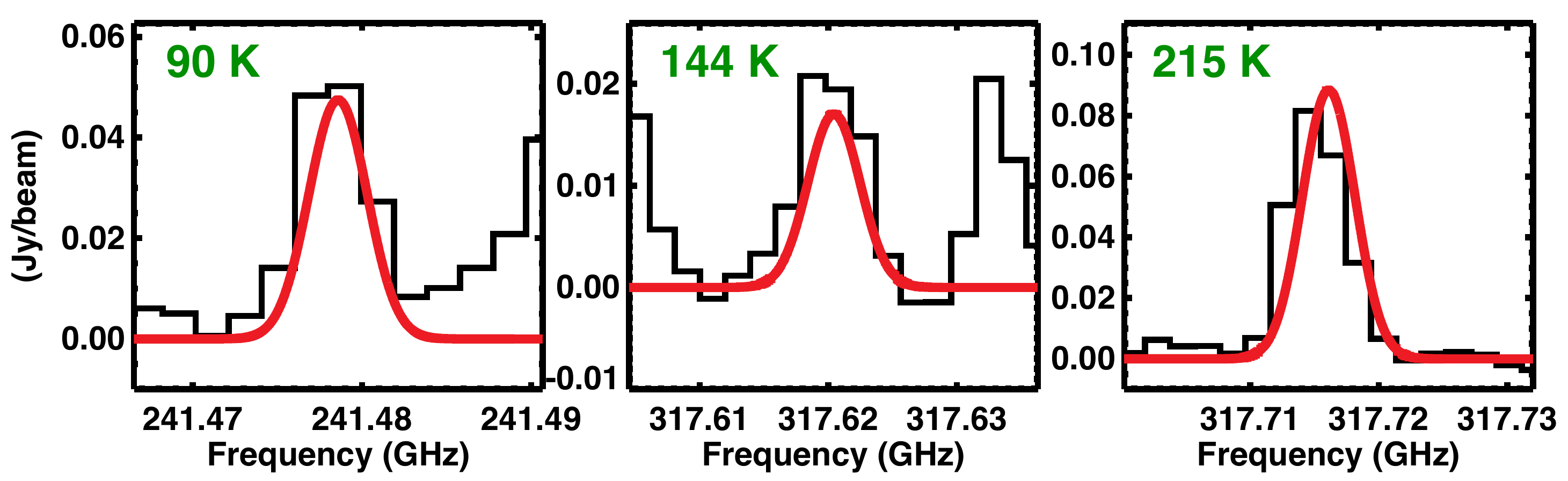}
\caption{Unblended lines of NH$_2$CN detected with PdBI towards IRAS2A. The best-fit model is shown in red. The $E_{\rm up}$ values are indicated in green in the upper left corner of each panel.}
\label{figure_detection_iras2a}
\end{center}
\end{figure}

Three bright and unblended lines of cyanamide are also detected towards NGC1333 IRAS2A (see Figure \ref{figure_detection_iras2a}). 
An excitation temperature of $\sim$130 K and a source size of $\sim$0.5$\arcsec$ were derived from the analysis of other complex organics towards this source by \citet{Coutens2015}. Based on these parameters, a LTE model with a column density of 2.5\,$\times$\,10$^{14}$ cm$^{-2}$ is in very good agreement with the observations. An excitation temperature of 300\,K does not properly reproduce the lines. No other line is missing in the spectral range covered by our data.
Based on this model, we can confirm the detection of this molecule by comparing our predictions with the list of lines observed in the spectral range 216.8--220.4 GHz covered by the CALYPSO program \citep{Maury2014}. Among the six brightest lines, five of them are in agreement with the presence of unidentified lines detected by \citet{Maury2014} (see Table \ref{table_U_line_IRAS2A}). The last one is blended with a NH$_2$CHO line. Maps confirm that the molecule is emitting within the warm inner regions of the source (see Figure \ref{maps_NH2CN}).

We also searched for carbodiimide, HNCNH, which is not detected with an upper limit of 3\,$\times$\,10$^{15}$ cm$^{-2}$ towards IRAS16293 B. Based on the non-detection of the HNCNH line at 223.7918 GHz, we derived a similar upper limit for IRAS2A.
The non-detection of carbodiimide is not really surprising since the upper limits are high and this isomer is known to be less stable than cyanamide.

\subsection{Analyses of the deuterated and $^{13}$C isotopologues of cyanamide}
\label{sect_isotopologues}

\begin{figure*}[ht!]
\begin{center}
\includegraphics[width=18cm]{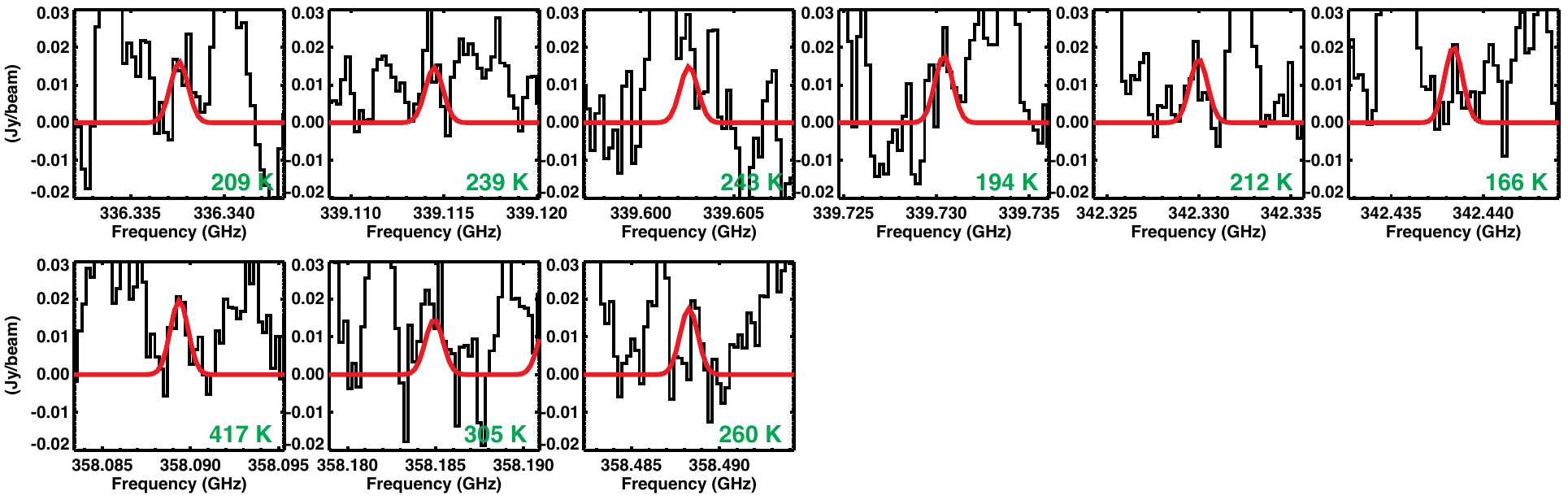}
\caption{Unblended lines of NHDCN detected with ALMA towards IRAS16293 B at the full-beam offset position. The best-fit model for $T_{\rm ex}$ = 300 K is in red. The $E_{\rm up}$ values are indicated in green in the bottom right corner of each panel.}
\label{figure_detection_NHDCN_iras16293}
\end{center}
\end{figure*}

\begin{figure}[ht!]
\begin{center}
\includegraphics[width=9cm]{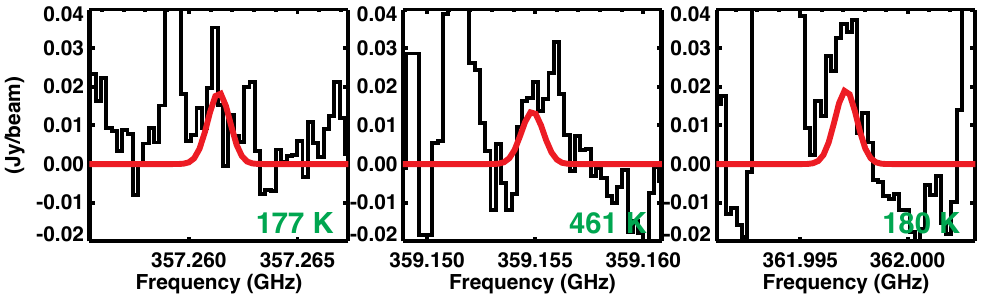}
\caption{Lines of NH$_2$$^{13}$CN detected with ALMA towards IRAS16293 B at the full-beam offset position. The best-fit model for $T_{\rm ex}$ = 300 K is in red. The $E_{\rm up}$ values are indicated in green in the bottom right corner of each panel.}
\label{figure_detection_NH213CN_iras16293}
\end{center}
\end{figure}

The $^{13}$C and deuterated isotopologues of NH$_2$CN were searched for towards both sources. 
Eight unblended lines of NHDCN are identified towards the full-beam offset position of IRAS16293 B (see Figure \ref{figure_detection_NHDCN_iras16293}). This marks the first detection of this isotopologue in the interstellar medium. Although the lines are faint, we can confirm that all the features are real after checking the spectra at the half-beam position (position where the lines are brighter). Additional lines are present but not included here, as they are blended with other species. A few other lines do not appear as bright as expected, due to the presence of absorptions produced by other species at the same frequency. A column density of 7\,$\times$\,10$^{13}$ cm$^{-2}$ is obtained for an excitation temperature of 300 K. 
The detection of NH$_2$$^{13}$CN is less straightforward. Three lines can be attributed to this isotopologue (see Figure \ref{figure_detection_NH213CN_iras16293}). One of them (361.997 GHz) may be blended with an unknown species since its flux is higher than predicted by the LTE model.
Based on these lines, a column density of 3\,$\times$\,10$^{13}$ cm$^{-2}$ is derived showing that the NH$_2$CN column density is underestimated due to the contribution of the absorptions. The D/H ratio of NH$_2$CN (corrected for statistics, i.e. divided by 2) is 1.7\%, assuming a standard $^{12}$C/$^{13}$C ratio of 68 \citep{Milam2005}. This value is very similar to the deuteration of formamide \citep[$\sim$2\%;][]{Coutens2016} and also within the range of the D/H ratios of other COMs \citep[$\sim$1--8\%,][]{Jorgensen2017,Persson2017}. In case of a lower $^{12}$C/$^{13}$C ratio of 30, which was found for a few COMs in this source \citep{Jorgensen2016,Jorgensen2017}, the D/H ratio of NH$_2$CN would be about 4\%.

For IRAS2A, an upper limit of 5\,$\times$\,10$^{13}$\,cm$^{-2}$ is derived for NHDCN, leading to a D/H ratio of $\leq$\,10\%. The $^{13}$C isotopologue also  presents an upper limit of 5\,$\times$\,10$^{13}$ cm$^{-2}$, which is equivalent to a $^{12}$C/$^{13}$C ratio of $\geq$\,5.

\subsection{Abundances of cyanamide}

Based on the analysis of the NH$_2$$^{13}$CN isotopologue and the lower limit of the H$_2$ column density derived by \citet{Jorgensen2016}, we get, for IRAS16293 B, an abundance of NH$_2$CN with respect to H$_2$ of $\lesssim$ 2\,$\times$\,10$^{-10}$ at 60 AU from source B. The NH$_2$CN/NH$_2$CHO abundance ratio is about 0.2. 

Using the H$_2$ column density derived by \citet[][5\,$\times$\,10$^{24}$ cm$^{-2}$]{Taquet2015} for IRAS2A, the abundance with respect to H$_2$ is about 5\,$\times$\,10$^{-11}$. 
A simple analysis of the most optically thin NH$_2$CHO lines covered by our data suggests a column density of about 1.2\,$\times$\,10$^{16}$\,cm$^{-2}$ assuming a similar excitation temperature of 130 K. This is in agreement with the value determined by \citet[][]{Taquet2015}. The NH$_2$CN/NH$_2$CHO ratio is consequently about 0.02, an order of magnitude lower than for IRAS16293 B. 

The NH$_2$CN/NH$_2$CHO ratios derived for the low-mass protostars are similar to the range of values found in Sgr B2 \citep[][see Table \ref{comp_model}]{Belloche2013}. The NH$_2$CN/NH$_2$CHO ratio in Orion KL seems to be higher ($\sim$0.4--1.5, \citealt{White2003}). This value is, however, more uncertain since it was obtained using only one line of NH$_2$CHO and assuming different excitation temperatures for the two molecules.

\begin{table*}[t!]
\begin{center}
\caption{Abundances of NH$_2$CN with respect to H$_2$ and NH$_2$CHO.}
\label{comp_model}
\vspace{-0.3cm}
\begin{tabular}{cccc}
\hline \hline
Source & ${\rm NH_2CN }/{\rm H_2}$ & ${\rm NH_2CN}/{\rm NH_2CHO}$ & Telescope \\
\hline
IRAS16293 B & $\lesssim$ 2\,$\times$\,10$^{-10}$ $^{(a)}$ &  $\sim$ 0.20 & ALMA \\
NGC1333 IRAS2A & $\sim$ 5\,$\times$\,10$^{-11}$ & $\sim$ 0.02 & PdBI \\
Sgr B2 (N) & -- & $\sim$ 0.02--0.04 & IRAM-30m \\
Sgr B2 (M) & -- & $\sim$ 0.15          & IRAM-30m \\
Orion KL & -- & $\sim$ 0.4--1.5        & JCMT \\
\hline
Low-density model & $\sim$ 3.7\,$\times$\,10$^{-10}$ & $\sim$ 1.1\,$\times$\,10$^{-3}$ \\
High-density model & $\sim$ 6.7\,$\times$\,10$^{-12}$ & $\sim$ 1.3\,$\times$\,10$^{-4}$\\
\hline
\end{tabular}
\vspace{-0.3cm}
\end{center}
\begin{center}
$^{(a)}$ \footnotesize{Based on the lower limit of 1.2\,$\times$\,10$^{25}$ cm$^{-2}$ for the H$_2$ abundance derived by \citet{Jorgensen2016}.}
\end{center}
\end{table*}%

\section{Discussion}
\label{sect_discussion}

The formation routes of NH$_2$CN have only been marginally explored.
According to the Kinetic Database for Astrochemistry\footnote{\url{http://kida.obs.u-bordeaux1.fr}} (KIDA, \citealt{Wakelam2012}), there are no known gas-phase mechanisms capable of its production. While the reaction CN + NH$_3$ $\rightarrow$ NH$_2$CN + H has been proposed \citep{Smith2004}, the theoretical study of \citet{Talbi2009} suggests that the production of NH$_2$CN involves large internal barriers, with HCN and NH$_2$ being the likely products. 
An experimental study by \citet{Blitz2009} confirms that this reaction proceeds exclusively to HCN + NH$_2$. 
Electronic recombination of NH$_2$CNH$^+$ may produce NH$_2$CN, but the only apparent way to form this ion is through protonation of NH$_2$CN itself. An alternative source of NH$_2$CN is thus required to explain our observations. 
Cyanamide could be formed on grain surfaces through the addition of NH$_2$ and CN radicals. The possible formation of formamide from the same precursor NH$_2$ \citep{Fedoseev2016} could explain the similarity of these two species in terms of deep absorption against the strong continuum and the similar deuteration of the two species towards IRAS16293 B.

To test this hypothesis, we ran a three-phase chemical kinetics model MAGICKAL \citep{Garrod2013}, modified with the grain-surface back-diffusion correction of \citet{Willis2017}. The model uses a network based on that of \citet{Belloche2017}, in which dissociative recombination of NH$_2$CNH$^+$ was assumed to produce NH$_2$CN in 5\% of cases. The reaction NH$_2$ + CN $\rightarrow$ NH$_2$CN was added to the grain/ice chemical network, and the gas-phase reaction between CN and NH$_3$ was adjusted per \citet{Talbi2009} and \citet{Blitz2009}. The physical model used here is very similar to that described by \citet{Belloche2017}, in which a cold collapse to high density is followed by warm up to 400~K; here, a final density $n_{\mathrm{H}}=6\times10^{10}$ cm$^{-3}$ was assumed to better represent the density structure of IRAS 16293 B \citep{Jorgensen2016}.

\begin{figure}[t!]
\begin{center}
\includegraphics[width=9cm]{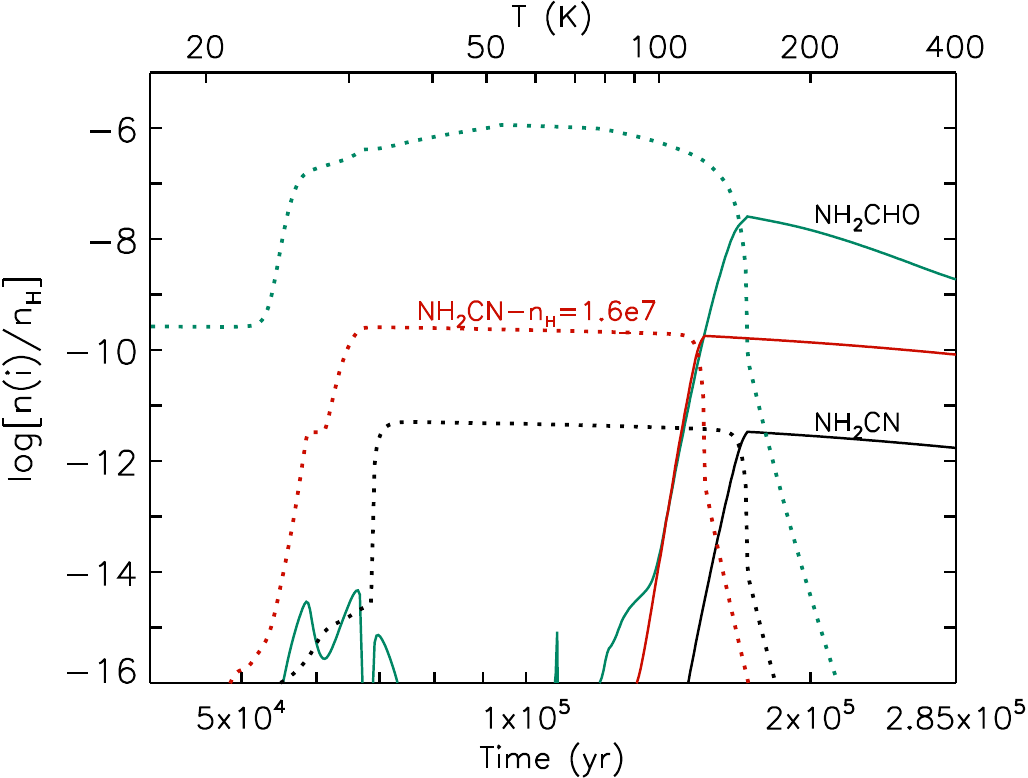} 
\caption{Chemical model abundances for the warm-up stage of a hot-core type model with a final collapse density of $n_{\mathrm{H}}$\,=\,6\,$\times$\,10$^{10}$ ~cm$^{-3}$ (high density model, black lines for NH$_2$CN and green lines for NH$_2$CHO). Solid lines denote gas-phase molecules; dotted lines indicate the same species on the grains. The red lines correspond to the abundance profiles of gas-phase and grain-surface NH$_2$CN for the lower density model, run at $n_{\mathrm{H}}$\,=\,1.6\,$\times$\,10$^{7}$ cm$^{-3}$.}
\label{fig_chemistry}
\end{center}
\end{figure}

The model results (for an intermediate warm-up timescale) are shown
in Figure \ref{fig_chemistry} for both NH$_2$CN and NH$_2$CHO. 
 NH$_2$CN is seen to be produced at a temperature of $\sim$30\,K on the grain surfaces, desorbing into the gas at higher temperatures.
The model underproduces the amount of gas-phase NH$_2$CN, showing a peak fractional abundance with respect to H$_2$ of $\sim$\,6.7\,$\times$\,10$^{-12}$ that is nevertheless well maintained to a temperature of 300\,K and beyond. 
 The low NH$_2$CN abundance in the gas-phase is caused primarily by underproduction on the dust grains;
at the high density used in the model, the rapid accretion of H and H$_2$ onto grain surfaces makes hydrogenation of the NH$_2$ and CN radicals
much more competitive with the reaction that produces NH$_2$CN. This competition becomes important for gas densities greater than $\sim$10$^9$ cm$^{-3}$.
We therefore also present a model with a lower final density of
$n_{\rm H}$ = 1.6\,$\times$\,10$^7$ cm$^{-3}$ (corresponding to the density of the envelope
between the two protostars in IRAS 16293, \citealt{Jacobsen2017}), intended to represent the
approximate conditions of the gas while at a temperature of 30\,K (see Figure \ref{fig_chemistry}). This model produces an
NH$_2$CN fractional abundance of 3.7\,$\times$\,10$^{-10}$, a value very close to the
detected value. However, the resulting NH$_2$CN:NH$_2$CHO peak abundance ratio of 0.0011 is
still lower than the observed values in both IRAS2A ($\sim$0.02), and
IRAS 16293 ($\sim$0.2). This may be due to the possible overproduction of
NH$_2$CHO, related to uncertainties in the efficiency of formation of
that molecule, which is still a matter of debate, particularly for the
gas-phase mechanism \citep[e.g.,][]{Barone2015,Song2016}; our
model assumes only a grain-surface/ice formation route.
The difficulty in reproducing the observed NH$_2$CN abundance at the high
densities determined for the source highlights the necessity for future
models of hot-core/corino chemistry to treat the rising density and
temperature in such cores concurrently, rather than as a two-stage process,
so that the gas densities are appropriate at the key temperatures at which
many molecules are formed.

In conclusion, the detection of cyanamide towards IRAS16293 B and IRAS2A indicates that this species can be formed early in solar-type protostars. If it survives during the star formation process until its incorporation into comets or asteroids, these objects could then deliver it to planets, which may enable the development of life. Search for this species in the coma of comets could shed further light on this possibility. Theoretical and experimental studies as well as more detailed chemical models are needed to confirm the formation of NH$_2$CN through the grain-surface pathway NH$_2$ + CN. It would also be interesting to investigate if this mechanism is sufficient to explain the large-scale emission of NH$_2$CN in galaxies.

\bibliographystyle{aa} 
\bibliography{Biblio} 

\begin{acknowledgements}
This paper makes use of the ALMA data ADS/JAO.ALMA$\#$2013.1.00278.S. ALMA is a partnership of ESO (representing
  its member states), NSF (USA) and NINS (Japan), together with NRC
  (Canada) and NSC and ASIAA (Taiwan), in cooperation with the Republic of
  Chile. The Joint ALMA Observatory is operated by ESO, AUI/NRAO and NAOJ.
  This work is based on observations carried out under project numbers V010, V05B, W00A, and X060 with the IRAM Plateau de Bure Interferometer. IRAM is supported by INSU/CNRS (France), MPG (Germany) and IGN (Spain). The research leading to these results has received funding from the European Commission Seventh Framework Programme (FP/2007-2013) under grant agreement No. 283393 (RadioNet3). A.C. postdoctoral grant is funded by the ERC Starting Grant 3DICE (grant agreement 336474). The group of J.K.J. acknowledges support from ERC Consolidator Grant "S4F" (grant agreement 646908). Research at the Centre for Star and Planet Formation is funded by the Danish National Research Foundation. The group of E.v.D. acknowledges ERC Advanced Grant "CHEMPLAN" (grant agreement 291141). MVP postdoctoral position is funded by the ERC consolidator grant 614264. MND acknowledges the financial support of the Center for Space and Habitability (CSH) Fellowship and the IAU Gruber Foundation Fellowship.
  \end{acknowledgements}

\appendix

\section{Tables with detected lines of NH$_2$CN and its isotopologues}

\begin{table}[h!]
\begin{center}
\caption{Detected lines of NH$_2$CN towards IRAS16293 B with ALMA}
\label{table_detected_16293}
\begin{tabular}{@{}ccccc@{}}
\hline \hline
 Transition  & Frequency & $E_{\rm up}$ & $A_{\rm ij}$ & $g_{\rm up}$ \\
(N, K$_{\rm a}$, K$_{\rm c}$, $\varv$)  & (MHz) & (K) & (s$^{-1}$)  \\
\hline
	17 2 15 1 -- 16 2 14 1 & 339238.0 & 274.3 & 3.9\,$\times$\,10$^{-3}$ & 105 \\
	17 0 17 1 -- 16 0 16 1 & 339450.6 & 218.0 & 4.04\,$\times$\,10$^{-3}$ & 105 \\
	17 0 17 0 -- 16 0 16 0 & 339710.9 & 146.8 & 4.15\,$\times$\,10$^{-3}$ & 35 \\
	17 2 15 0 -- 16 2 14 0 & 339892.9 & 204.8 & 4.10\,$\times$\,10$^{-3}$ & 35 \\
	18 1 18 0 -- 17 1 17 0 & 357404.4 & 177.5 & 4.82\,$\times$\,10$^{-3}$ & 111 \\
	18 2 16 1 -- 17 2 15 1 & 359203.0 & 291.6 & 4.65\,$\times$\,10$^{-3}$ & 111 \\
	18 2 16 0 -- 17 2 15 0 & 359892.2 & 222.0 & 4.88\,$\times$\,10$^{-3}$ & 37 \\
	18 3 16 0 -- 17 3 15 0 & 360114.0 & 294.5 & 4.87\,$\times$\,10$^{-3}$ & 111 \\
	18 3 15 0 -- 17 3 14 0 & 360127.4 & 294.6 & 4.87\,$\times$\,10$^{-3}$ & 111 \\
	18 1 17 1 -- 17 1 16 1 & 361717.6 & 250.4 & 4.88\,$\times$\,10$^{-3}$ & 37 \\
	18 1 17 0 -- 17 1 16 0 & 362143.5 & 179.6 & 5.02\,$\times$\,10$^{-3}$ & 111 \\
\hline
\end{tabular}
\end{center}
\end{table}%

\begin{table}[h!]
\begin{center}
\caption{Detected lines of NH$_2$CN towards IRAS2A with PdBI}
\label{table_detected_IRAS2A}
\begin{tabular}{@{}ccccc@{}}
\hline \hline
 Transition & Frequency & $E_{\rm up}$ & $A_{\rm ij}$ & $g_{\rm up}$ \\
 (N, K$_{\rm a}$, K$_{\rm c}$, $\varv$) & (MHz) & (K) & (s$^{-1}$)  \\
\hline
12 1 11 0 -- 11 1 10 0 & 241478.6 & ~89.8 & 1.46\,$\times$\,10$^{-3}$ & 75 \\
16 1 16 1 -- 15 1 15 1 & 317620.4 & 215.0 & 3.29\,$\times$\,10$^{-3}$ & 33 \\
16 1 16 0 -- 15 1 15 0 & 317716.1 & 144.1 & 3.37\,$\times$\,10$^{-3}$ & 99 \\
\hline
\end{tabular}
\end{center}
\end{table}%

\begin{table}[h!]
\begin{center}
\caption{Detected lines of NHDCN towards IRAS16293 B with ALMA}
\label{table_detected_16293B_NHDCN}
\begin{tabular}{@{}ccccc@{}}
\hline \hline
 Transition & Frequency & $E_{\rm up}$ & $A_{\rm ij}$ & $g_{\rm up}$ \\
 (N, K$_{\rm a}$, K$_{\rm c}$, $\varv$) & (MHz) & (K) & (s$^{-1}$)  \\
\hline
 18  1 18 1 -- 17 1 17 1 & 336337.6 & 209.2 & 3.93\,$\times$\,10$^{-3}$ & 37 \\
 18 2 17 1 -- 17 2 16 1 & 339114.5 & 239.4 & 3.94\,$\times$\,10$^{-3}$ & 37 \\
 18 3 16 0 -- 17 3 15 0 & 339602.6 & 243.2 & 4.07\,$\times$\,10$^{-3}$ & 37 \\
 18 2 16 0 -- 17 2 15 0 & 339730.4 & 194.1 & 4.10\,$\times$\,10$^{-3}$ & 37 \\
18 1 17 1 -- 17 1 16 1 & 342330.0 & 211.9 & 4.14\,$\times$\,10$^{-3}$ & 37 \\
18 1 17 0 -- 17 1 16 0 & 342438.4 & 166.0 & 4.24\,$\times$\,10$^{-3}$ & 37 \\
19 5 15 0 -- 18 5 14 0 & 358089.4 & 416.9 & 4.54\,$\times$\,10$^{-3}$ & 39 \\
19 5 14 0 -- 18 5 13 0 & 358089.4 & 416.9 & 4.54\,$\times$\,10$^{-3}$ & 39 \\
19 3 17 1 -- 18 3 16 1 & 358184.9 & 304.9 & 4.63\,$\times$\,10$^{-3}$ & 39 \\
 19 3 16 0 -- 18 3 15 0 & 358488.3 & 260.4 &  4.80\,$\times$\,10$^{-3}$ & 39 \\ 
 \hline
\end{tabular}
\end{center}
\end{table}%

\begin{table}[h!]
\begin{center}
\caption{Detected lines of NH$_2$$^{13}$CN towards IRAS16293 B with ALMA}
\label{table_detected_16293B_NH213CN}
\begin{tabular}{@{}ccccc@{}}
\hline \hline
 Transition & Frequency & $E_{\rm up}$ & $A_{\rm ij}$ & $g_{\rm up}$ \\
 (N, K$_{\rm a}$, K$_{\rm c}$, $\varv$) & (MHz) & (K) & (s$^{-1}$)  \\
 \hline
 18 1 18 0 -- 17 1 17 0 & 357261.4 & 177.4  &  4.82\,$\times$\,10$^{-3}$ & 111 \\
 18 4 15 1 -- 17 4 14 1 & 359154.9 & 460.9  & 4.55\,$\times$\,10$^{-3}$ & 111 \\
 18 4 14 1 -- 17 4 13 1 & 359154.9 & 460.9  & 4.55\,$\times$\,10$^{-3}$ & 111 \\
 18 1 17 0 -- 17 1 16 0 & 361997.1 & 179.6 & 5.01\,$\times$\,10$^{-3}$ & 111 \\
\hline
\end{tabular}
\end{center}
\end{table}%

\begin{table}[h!]
\begin{center}
\caption{Unidentified lines from \citet{Maury2014} (their Table 1) that can be assigned to NH$_2$CN. These data have a spectral resolution of 3.9 MHz ($\sim$ 5.3 km\,s$^{-1}$).}
\label{table_U_line_IRAS2A}
\begin{tabular}{cc}
\hline \hline
 Rest frequency & Frequency of the U-line \\ 
 (MHz) & (MHz) \\
 \hline
218461.8 & No U-line  but blending \\
& with NH$_2$CHO at 218459 MHz \\
219441.6 & 219441 \\
219474.0 & 219474 \\
219719.7 & 219719 \\
219893.8 & 219892 \\
220126.6 & 220126 \\
220127.9 & 220126 \\
\hline
\end{tabular}
\end{center}

\end{table}%

\section{Spectra of NH$_2$CN observed with ALMA at different positions towards IRAS16293~B }

\begin{figure*}[h!]
\begin{center}
\includegraphics[width=16cm]{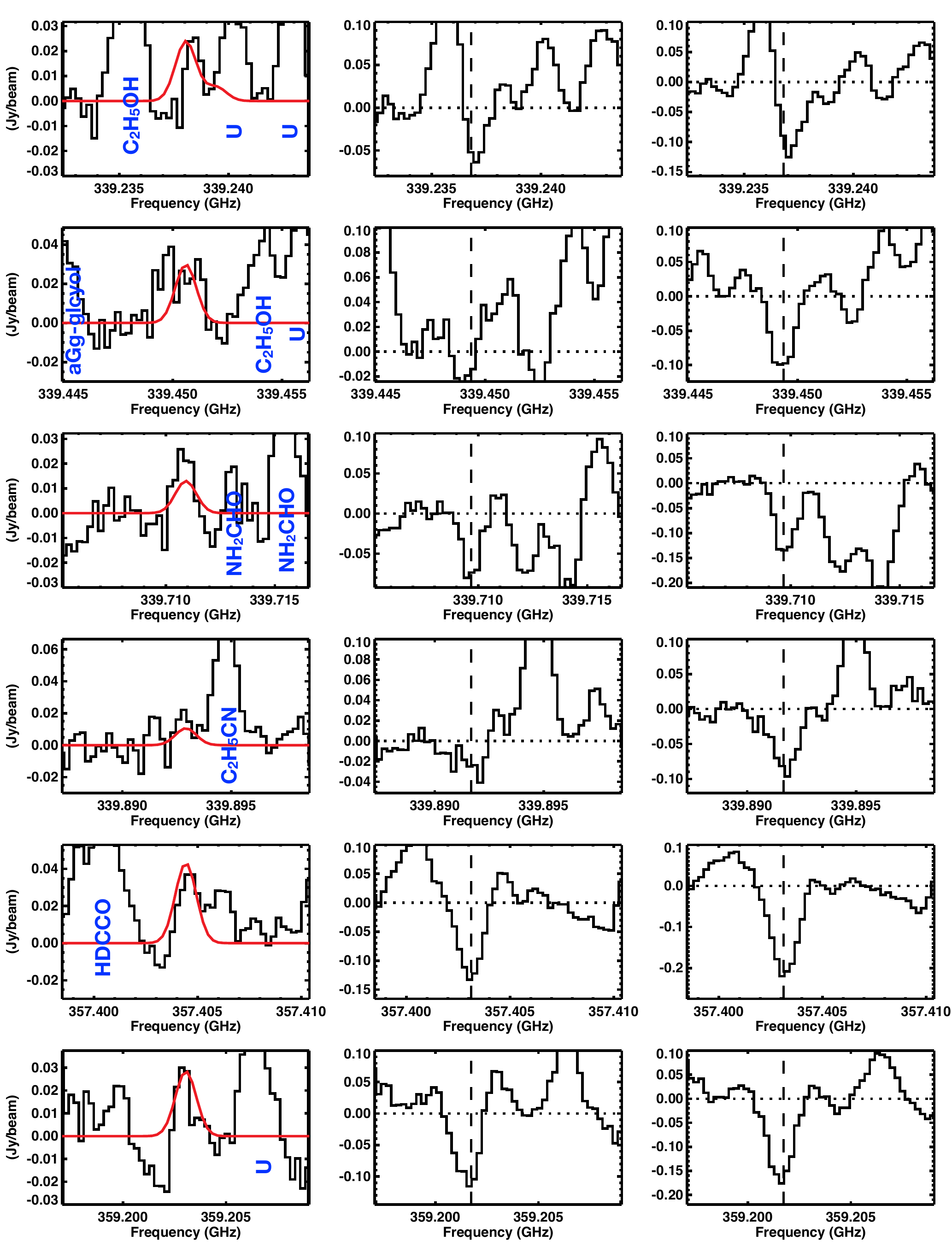}
\caption{Spectra of the unblended lines of NH$_2$CN detected with ALMA towards IRAS16293 B at the full-beam offset position (left), at the half-beam offset position (middle) and at the peak continuum position (right). The best-fit model for $T_{\rm ex}$ = 300 K at the full-beam offset position is shown in red on the left panels. The identification of the other lines is indicated in blue. The dashed line on the middle and right panels indicates the average velocity of the absorptions, 3.8 km\,s$^{-1}$.The dotted line shows the level 0.}
\label{figure_absorption}
\end{center}
\end{figure*}
\begin{figure*}[h!]
\begin{center}
\ContinuedFloat
\includegraphics[width=16cm]{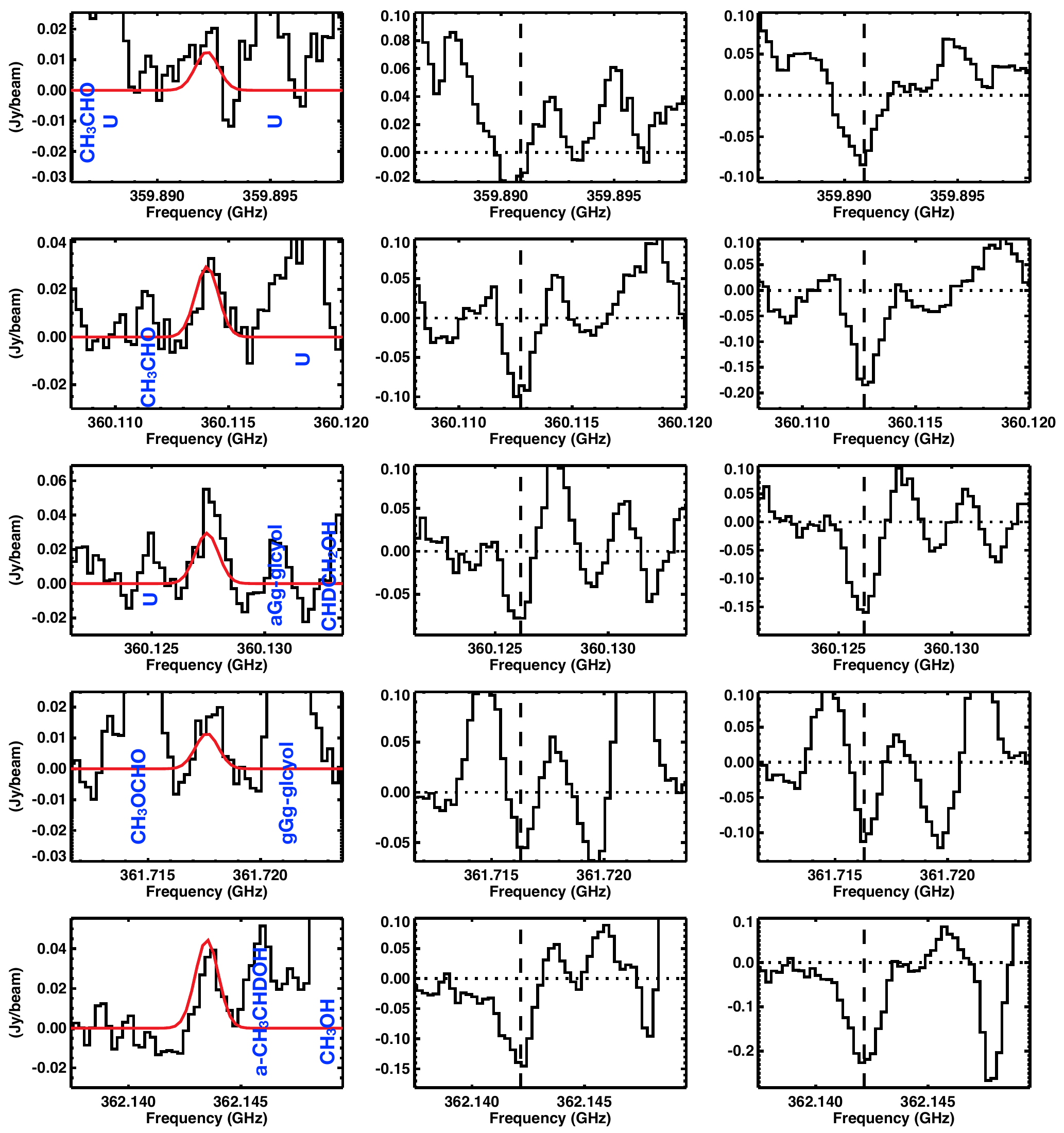}
\caption{Spectra of the unblended lines of NH$_2$CN detected with ALMA towards IRAS16293 B at the full-beam offset position (left), at the half-beam offset position (middle) and at the peak continuum position (right). The best-fit model for $T_{\rm ex}$ = 300 K at the full-beam offset position is shown in red on the left panels. The identification of the other lines is indicated in blue. The dashed line on the middle and right panels indicates the average velocity of the absorptions, 3.8 km\,s$^{-1}$.The dotted line shows the level 0.}
\end{center}
\end{figure*}

\end{document}